\documentclass[11pt]{article}
\usepackage[cp866]{inputenc}

\newcommand{\const}{{\rm\, const}}

\makeatletter

\renewcommand{\@oddhead}{\hfil Alexander Shatskiy $\qquad\qquad$  "Dynamics of Phantom Matter"}

\input epsf

\begin{document}

\begin{center}{\bf  {\Large Dynamics of Phantom Matter} \\
\quad \\
Alexander Shatskiy *\\
${}^*$ AstroSpaceCenter, Lebedev Institute of Physics, Russian
Academy of Sciences}
\end{center}


DOI: 10.1134/S1063776107050081

$\quad$ \\

{\bf\large Abstract}\\
A spherically symmetric evolution model of self-gravitating matter
with the equation of state ${p=-(1+\delta)\varepsilon}$ (where
${\delta=\const}$) is considered. The equations of the model are
written in the frame of reference comoving with matter. A
criterion for the existence and formation of a horizon is defined.
Part of the Einstein equations is integrated analytically. The
initial conditions and the constraints imposed on these conditions
in the presence of a horizon are determined. For small $\delta$ an
analytic solution to spherically symmetric time-dependent Einstein
equations is obtained. Conditions are determined under which the
dynamics of matter changes from collapse to expansion.
Characteristic times of the evolution of the system are evaluated.
It is proved that the accretion of phantom matter (for
${\delta>0}$) onto a black hole leads to the decreases of the
horizon radius of the black hole (i.e., the black hole is
"dissolved").

\section{INTRODUCTION}
\label{s1}

Matter that violates the zero energy condition (NEC) is said to be
phantom matter~\cite{10}. For matter that violates this condition,
the sum of the energy density ${\varepsilon}$ and pressure $p$,
${(\varepsilon +p)}$ is less than zero. In the present paper, we
assume that the energy density $\varepsilon$ of phantom matter is
positive (in the co-moving frame of reference). The sum
${(\varepsilon +p)}$ is proportional to the energy density in the
frame of reference co-moving with a photon~\cite{5}. Of course,
such a frame of reference is senseless; only the asymptotic limit
for a frame of reference whose velocity approaches the velocity of
light makes sense.

Phantom matter has a number of exotic properties; this fact gives
a motivation for its comprehensive study.

First, only in this matter "wormholes" can exist (see, for
example,~\cite{9} or \cite{4}).

Second, there is a conjecture~\cite{3} that phantom matter
possesses a unique property to dissolve black holes in it. Until
now, this has been proved only for a non-self-consistent solution
(when the effect of the gravitation of this matter can be
neglected compared with the gravitation of ordinary matter). The
complete dissolution of a black hole in phantom matter would
require revision of the definition and the description of the term
"event horizon"\, (see Section~\ref{s3}).

Third, the state of the art in observational cosmology allows one
to speak of possible prevalence of phantom matter in the universe
(according to the latest measurements of the acceleration of the
expanding universe). In addition, a real example of phantom matter
is given by the quantum field of the Casimir effect (between close
conducting plates~\cite{8}). Thus, the study of such matter is of
practical interest~\cite{6},\cite{9}.

Fourth, there is an analytic solution to the Einstein equations
for the cosmological equation of state ${\varepsilon +p =0}$,
therefore, small deviations from this exact equality are possible
toward the equation of state corresponding to the case of phantom
matter, and it becomes possible to find a first-approximation
solution in a small parameter.

The analytical solution thus obtained removes the above-mentioned
contradictions and questions.

\section{EQUATIONS OF MOTION}
\label{s2}

Let us choose a frame of reference co-moving with matter (see, for
example~\cite{1}). It is convenient to choose the metric tensor in
the spherically symmetric case as\footnote{We use a system of
units in which $c=1$  and $G= 1$.}:
\begin{equation}
ds^2=e^\nu dt^2 - e^\lambda dR^2 - e^\mu (d\theta^2
+\sin^2\theta\, d\varphi^2). \label{1-1}\end{equation} Here, the
metric component $e^\mu =r^2$ determines the area of the sphere
around the center of the system ($4\pi r^2$); the metric
components $r$, $\nu$ and $\lambda$ are functions of the
coordinates $R$ and $t$.

The Einstein equations corresponding to metric (\ref{1-1}) can be
expressed as\footnote{These equations are derived, for example,
in~\cite{1} (Problem 5 in \S 100).}:
\begin{equation}
\begin{array}{l}
8\pi\varepsilon = -e^{-\lambda}\left( \mu'' +{3\over 4}(\mu')^2 -
{1\over 2}\mu'\lambda' \right) + e^{-\nu} \left({1\over
2}\dot\mu\dot\lambda +{1\over 4}(\dot\mu)^2 \right) +e^{-\mu}\, ,
\end{array}\label{1-2}\end{equation}
\begin{equation}
\begin{array}{l}
8\pi p_\parallel = -e^{-\nu}\left( \ddot\mu + {3\over
4}(\dot\mu)^2 - {1\over 2}\dot\mu\dot\nu \right) + e^{-\lambda}
\left({1\over 2}\mu'\nu' +{1\over 4}(\mu')^2 \right)- e^{-\mu}\, ,
\end{array}\label{1-3}\end{equation}
\begin{equation}
\begin{array}{l}
8\pi p_\perp = e^{-\lambda}\left({1\over 2}\nu'' + {1\over
4}(\nu')^2 + {1\over 2}\mu'' +{1\over 4}(\mu')^2 -
{1\over 4}\mu'\lambda' - {1\over 4}\nu'\lambda' + {1\over 4}\mu'\nu' \right) +\\
+ e^{-\nu}\left({1\over 4}\dot\lambda\dot\nu +{1\over
4}\dot\mu\dot\nu - {1\over 4}\dot\lambda\dot\mu - {1\over
2}\ddot\lambda - {1\over 4}(\dot\lambda)^2
 -{1\over 2}\ddot\mu - {1\over 4}(\dot\mu)^2 \right)\, ,
\end{array}\label{1-4}\end{equation}
\begin{equation}
\begin{array}{l}
2\dot\mu' +\dot\mu\mu' -\dot\lambda\mu' - \nu'\dot\mu =0\, .
\end{array}\label{1-5}\end{equation}
where ${p_\parallel =-T^R_R}$ is longitudinal pressure,
${p_\perp=-T^\theta_\theta=-T^\varphi_\varphi}$ is transverse
pressure, the prime denotes the derivative with respect to $R$,
the dot denotes the derivative with respect to $t$, and $T^k_i$
are the components of the energy-momentum tensor.

If pressure is isotropic, ${p_\parallel=p_\perp=p}$, then these
equations yield two useful relations, which can also be obtained
directly from the formula ${T^k_{i;k}=0}$ (the Bianchi
identities):
\begin{equation}
\begin{array}{l}
\dot\lambda +2\dot\mu = -2\dot\varepsilon /(p+\varepsilon) \,
,\quad \nu' =-2p'/(p+\varepsilon) \, .
\end{array}\label{1-6}\end{equation}
Let us write out the equation of state of the matter:
\begin{equation}
\begin{array}{l}
p= p_\perp =p_\parallel = w\varepsilon\, ,\quad w\equiv
-(1+\delta) \, .
\end{array}\label{1-7}\end{equation}
where $w$ is a constant parameter that determines this equation.
When $\delta =0$, Eqs. (\ref{1-6}) clearly show that
${-p=\varepsilon =\const}$, this implies the well-known
solution~\cite{Wym} with a $\Lambda$-term:
\begin{equation}
\begin{array}{l}
\varepsilon =\varepsilon_{_\Lambda}>0\, ,\quad
\nu_{_\Lambda}=-\lambda_{_\Lambda}=\ln (1- r_g/R-\Lambda R^2) \,
.
\end{array}\label{1-8}\end{equation}
In this expression, the constant ${r_g}$ corresponds to the
gravitational radius in the presence of a horizon, while the
constant ${\varepsilon_{_\Lambda}\equiv 3\Lambda /(8\pi)}$ is the
energy density of the cosmological $\Lambda$-term.

Due to the presence of the $\Lambda$-term, the metric becomes
unphysical for large $R$ (due to the external horizon). Therefore,
here one should define the maximum possible radius ${R_\infty}$:
\begin{equation}
\Lambda\, R^2_\infty =1\, . \label{1-8-1}\end{equation} In
addition, the asymptotics of the metric for large $R$ must be
Galilean:
\begin{equation}
\nu_\infty\to 0\, . \label{1-8-0}\end{equation} Below, when
speaking of physical infinity, we will assume that the boundaries
of the metagalaxy are not reached, i.e., ${R<R_\infty}$.

\section{CRITERION FOR THE EXISTENCE
OF AN APPARENT HORIZON} \label{s3}

The radius of the event horizon depends on the entire future
evolution of matter~\cite{9}. This is inconvenient for describing
real black holes from both mathematical and practical points of
view because the definition of the event horizon is not local.
Usually, when speaking of a black hole, we assume that there
exists an apparent horizon and that the event horizon will be
formed at the end of evolution (it will be clear from the sequel
that this does not always correspond to reality).

Let us give a local definition of the apparent horizon $r_h$
~\cite{9}. We prove that the apparent horizon in a spherically
symmetric system is formed at the moment when an incident particle
with nonzero rest mass reaches the velocity of light with respect
to the surfaces ${r=\const}$ at the same point at which the
radiating particle is located at this moment.

Now suppose that we are on a particle of matter with coordinate
${r_{\infty}}$ and, from a large radius ${R_{\infty}}$, follow up
a particle of matter with coordinate ${R_p}$ that radiates light
while passing through spheres of radius ${r_{(R_p,t)}}$. A
criterion that a particle has not reached the horizon is the fact
that we can still see light emitted by this particle; i.e., light
intersects surfaces with radii ${r=\const}$. Hence, a criterion
that a particle reaches the horizon is the event that the
propagating light cannot intersect surfaces with radii ${r>
r_{(R_p,t)}}$. Let us express this criterion mathematically.

\begin{figure}[t]
\centering \epsfbox[20 300 500 600]{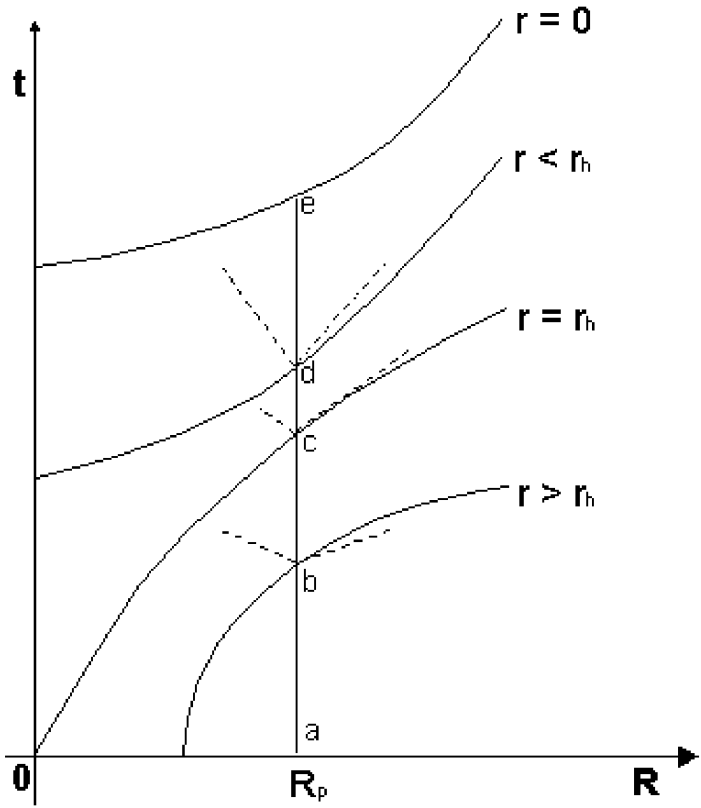}

\caption{The straight line ${a-b-c-d-e}$ is the world line of an
observer in the coordinates of the co-moving frame of reference.
The curves represent the graphs of ${t_{(R)}}$ for various
surfaces ${r=\const}$. The dashed lines are light cones within
which light can propagate. At the point $b$, the light cone
intersects the surfaces ${r=\const}$; hence, there is no horizon
at this point. At the point $d$, the light cone passes above the
curve ${r=\const}$; hence, it intersects only the surfaces with
${r< r_{(R_p,t)}}$; therefore, the point $d$ is below the
horizon.}

\label{R1} \end{figure}

In Fig.~\ref{R1}, the straight vertical line ${a-b-c-d-e}$ denotes
the world line of the particle $R_p$ in the coordinates of the
co-moving frame of reference ${(R,t)}$ from the moment of rest
($a$) to the center of the system ($e$), where ${r=0}$. The solid
curves passing through the points $e$, $d$, $c$, and $b$ represent
the lines of constant values of ${r_{(R,t)}}$ for ${r=0}$,
${r<r_h}$, ${r=r_h}$ and ${r>r_h}$, respectively. The dashed lines
with vertices at these points indicate cones within which light
emitted by particle $R_p$ can propagate. Therefore, according to
the above criterion, the horizon exists at the point where the
cone is tangent to the line ${r=\const}$; in Fig.~\ref{R1}, this
line corresponds to the value ${r=r_h}$ (it passes through the
point $c$).

 The condition of constancy of the radius ${r_{(R,t)}}$ has the form:
\begin{equation}
dr =0= \dot r\, dt + r'\, dR \, . \label{6-1}\end{equation} Hence,
the tangent of the slope of the curve ${r=\const}$ with respect to
the axis $R$ is
\begin{equation}
\left. {dt\over dR}\right|_{r=\const}=-{r'\over \dot r} \, .
\label{6-1-2}\end{equation} For the light cone, we have (by
definition) ${ds^2=0}$, which, with regard to (\ref{1-1}), yields
\begin{equation}
\left. {dt\over dR}\right|_{ds=0}=\sqrt{e^{\lambda-\nu}} \, .
\label{6-2} \end{equation} Then, a criterion for the absence of a
horizon is given by the condition:
\begin{equation}
\left. {dt\over dR}\right|_{ds=0}<\left. {dt\over
dR}\right|_{r=\const} \, . \label{6-3} \end{equation} Substituting
(\ref{6-1-2}) and (\ref{6-2}) into this expression and introducing
the physical velocity $V$ of the particle of matter, we rewrite
this criterion as
\begin{equation}
V^2 \equiv e^{\lambda -\nu}\left( \dot r/r'\right)^2 <1 \, .
\label{6-4} \end{equation} Introduce an element of proper
(physical) time
\begin{equation}
d\tau\equiv\exp (\nu /2)dt \, . \label{6-6} \end{equation} a
physical longitudinal distance $l$ and an element of physical
length $dl$:
\begin{equation}
l_{(R,t)}\equiv \int\limits_{_0}^R e^{\lambda /2}{}_{(R,t)}\, dR
\, ,\quad dl\equiv\exp (\lambda /2)\, dR \, . \label{6-7}
\end{equation} Then, the velocity of motion of matter with respect
to the surfaces ${r=\const}$ is given by
\begin{equation}
V = \left. {dl\over d\tau }\right|_{r=\const} = \sqrt{e^{\lambda
-\nu }} \left. {dR\over dt}\right|_{r=\const}= \sqrt{e^{\lambda
-\nu}}{\dot r\over r'} \, . \label{6-5} \end{equation} This proves
the assertion\footnote{This assertion is inapplicable at the mouth
of a wormhole (if any) because ${e^{-\lambda}}$ identically
vanishes there. Therefore, the existence condition of a horizon
near the mouth is determined asymptotically: ${V^2\to 1}$ at
${r\to r_h}$.}
 that the apparent horizon is formed at the moment when matter reaches the
velocity ${|V|=1}$ with respect to the surface ${r=\const}$.

\section{GENERAL INTEGRALS OF MOTION}
\label{s10}

Integrating the first equation in (\ref{1-6}) with regard to the
initial conditions, we obtain
\begin{equation}
e^\lambda = e^{\lambda_{_0}} \left({R\over r}\right)^4
\left({\varepsilon\over \varepsilon_{_0}}\right)^{2/\delta} \, .
\label{3-3-0}\end{equation}

The integration of the second equation in (\ref{1-6}) yields
\begin{equation}
\nu =-2{1+\delta\over\delta}\ln (\varepsilon
/\varepsilon_{_\Lambda}) + c_{_0}(t) \, ,
\label{3-3}\end{equation} By an admissible transformation of time
${ c_{_0}(t)}$, we can set the function ${ c_{_0}(t)}$ equal to
zero. Then,
\begin{equation}
\begin{array}{l}
\nu =-2{1+\delta\over\delta}\ln (\varepsilon
/\varepsilon_{_\Lambda})
\end{array}\label{3-2-1}\end{equation}
Then, condition (\ref{1-8-0}) is fulfilled automatically as
${R\to\infty}$.

Let us rewrite Eq. (\ref{1-5}) as
\begin{equation}
2{\dot r}' =r' \dot\lambda +\dot r \nu'  \, .
\label{1-5-2}\end{equation} Determine ${\dot\lambda}$ from this
equation and substitute it into Eq. (\ref{1-2}). We obtain
\begin{equation}
8\pi\varepsilon r^2={1\over r'}{d\over dR}\left(r- r {r'}^2
e^{-\lambda}+ r {\dot r}^2 e^{-\nu}\right)\, .
\label{9-3}\end{equation} Let us introduce a new quantity that has
the meaning of mass:
\begin{equation}
M_{(R,t)}\equiv\int\limits_{_0}^R 4\pi\varepsilon r^2 r'\, dR \, ,
\label{9-3-0}\end{equation} Upon integrating Eq. (\ref{9-3}), we
can obtain two equations
\begin{equation}
2M = r \left(1 - {r'}^2 e^{-\lambda}+ {\dot r}^2 e^{-\nu}\right)\,
,\quad V^2=1-{e^{\lambda}\over {r'}^2}\cdot \left( 1-{2M\over
r}\right) \, . \label{9-4}\end{equation} Hence, provided that
${\left. {r'}^2 e^{-\lambda}\right|_{r=r_h}\ne 0}$, the horizon
radius is
\begin{equation}
r_h =2M_{(R_h,t_h)} \, . \label{9-5}\end{equation} Note that Eq.
(\ref{1-3}) coincides with Eq. (\ref{1-2}) if we replace the time
derivative by the derivative with respect to the coordinate (and
vice versa), make the change ${\lambda\leftrightarrow\nu}$, and
change the signs in the last terms of these equations. Then, by
analogy, Eq. (\ref{9-3}) can be rewritten as
\begin{equation}
-8\pi p_\parallel r^2={1\over\dot r}{d\over dt}\left(r- r {r'}^2
e^{-\lambda} + r {\dot r}^2 e^{-\nu}\right) \, .
\label{9-3-2}\end{equation} Integrating with respect to time and
taking into account (\ref{9-4}) and the initial conditions, we
obtain
\begin{equation}
2M=2M_{_0} - \int\limits_{_0}^t 8\pi p_\parallel r^2 \dot r \,dt\,
, \label{9-7}\end{equation} Here,
\begin{equation}
M_{_0}\equiv M_{(R,0)}\, , \quad S_{(R)}\equiv 2M_{_0}/R \, .
\label{9-8}\end{equation} Formulas (\ref{9-3-0}) and (\ref{9-4})
yield an expression for the initial mass in the presence of a
horizon:
\begin{equation}
2M_{_0}\equiv r_{1}+\int\limits_{r_{1}}^R 8\pi\varepsilon_{_0} R^2
\, dR \, , \label{9-3-1}\end{equation} where ${r_{_1}}$ is the
initial radius of the horizon, which is defined by
${r_{_1}=r_g+\Lambda r_{_1}^3\approx r_g}$. Formula (\ref{9-7})
determines a shift of the horizon in matter due to the work of
pressure forces on the internal layers of matter (when matter is
collapsed to the center). In~\cite{2}, formula (\ref{9-7}) was
obtained for a particular case of  ${w=\const}$.

\section{INITIAL CONDITIONS IN THE ABSENCE OF A HORIZON}
\label{s4}

Let us specify the initial conditions of the problem for a model
in which there is no horizon at the initial moment. Using an
admissible transformation ${R=R_{(\tilde R)}}$, we can specify:
\begin{equation}
r_{_0}=R \quad\Longrightarrow\quad  r'_{_0}=1 \, .
\label{3-1-00}\end{equation} at the initial moment. In addition,
we can specify the condition that matter is at rest at the initial
moment:
\begin{equation}
\dot\mu_{_0} =0\, ,\quad V_{_0}=0\, ,\quad \dot\lambda_{_0} =0\,
,\quad \dot\nu_{_0} =0   \, . \label{3-1-0}\end{equation} Then,
Eqs. (\ref{9-4}) yield
\begin{equation}
\begin{array}{l}
\lambda_{_0}=-\ln (1-S)\, ,
\end{array}\label{1-9}\end{equation}
where
$$
S\equiv 2M_{_0}/R \, ,
$$
this result does not depend on the choice of $\delta$.

\section{INITIAL CONDITIONS IN THE PRESENCE OF A HORIZON}
\label{s5}

In this model, it is assumed that a horizon may exist at the
initial moment; this assumption also concerns a purely vacuum
solution with a $\Lambda$-term. Therefore, the initial conditions
near the horizon deserve special mentioning.

The initial condition (\ref{3-1-00}) is retained. With regard to
(\ref{6-4}), on the one hand, we have
$$
V^2_{_0}{}_{(R)} = (\dot r_{_0})^2\cdot\left. \exp (\lambda_{_0}
-\nu_{_0})\right|_{r_h} =1\, ,
$$
while, on the other hand,
$$
\dot r_{_0} =0\, ,\quad \left. e^{\lambda_{_0}}\right|_{r_h}
\to\infty \, .
$$
Therefore, we obtain indeterminacy of the form ${0\cdot\infty}$ on
the horizon. Removing this indeterminacy, we obtain a physical
distribution for velocity $V_{_0}$ at the initial moment:
\begin{equation}
|V_{_0}{}_{(R)}| = \left[
\begin{array}{l}
0 \quad (R\ne r_h) \, ,\\
1 \quad (R=r_h) \, .
\end{array}
\right. \label{9-1}\end{equation} i.e., the initial distribution
of velocity represents a function with a "punctured point."

To validate formula (\ref{9-1}), we modify the initial conditions:
we assume that ${\dot r_{_0} \ne 0}$. Rewrite the second
expression in (\ref{9-4}) as
\begin{equation}
e^{-\lambda_{_0}}=1+\alpha -S \, , \label{9-1-2}\end{equation}
where ${\alpha\equiv {\dot r_{_0}}^2 e^{-\nu_{_0}}}$ is determined
by the initial distribution of the velocity of matter.

Hence we obtain the following expression for the initial
distribution of velocity:
\begin{equation}
V_{_0}^2 = {\alpha \over 1+\alpha -S} \, .
 \label{9-6}\end{equation}
As ${\alpha\to 0}$, formula (\ref{9-6}) turns into (\ref{9-1}).

\section{ANALYTICAL SOLUTION}
\label{s6-0}

Denote
\begin{equation}
y\equiv \ln (\varepsilon /\varepsilon_{_\Lambda}) \, .
\label{12-1}\end{equation} As ${\delta\to 0}$, we have
${\varepsilon\to\varepsilon_{_\Lambda}}$; therefore, ${y\to 0}$ as
${\delta\to 0}$.

Substituting ${\dot\lambda}$ and ${\nu'}$ from Eqs. (\ref{1-6})
into Eq. (\ref{1-5-2}), we obtain
\begin{equation}
\delta\left( {\dot r'\over \dot r r'} +{2\over r} \right) = {\dot
y\over \dot r} - (1+\delta){y'\over r'} \, .
\label{12-2}\end{equation} Let us pass from the old coordinates
${(R,\, t)}$ to new coordinates ${(r,\, r_h)}$, which are
functions of the old coordinates. Applying a trivial mathematical
transformation, we obtain
\begin{equation}
{\dot y\over \dot r}-{y'\over r'}={\partial y\over\partial r_h}
\left({\dot r_h\over \dot r}-{{r_h}'\over r'} \right) .
\label{12-2-2}\end{equation} In Section~\ref{s8}, we will show
that, for small $\delta$, the variation of ${r_h}$ is also small
(and proportional to $\delta$). Taking into account that ${y\to
0}$ as ${\delta\to 0}$, we can see that the right-hand side of
(\ref{12-2-2}) can be neglected as ${\delta\to 0}$. Then, Eq.
(\ref{12-2}) is rewritten as
\begin{equation}
\delta\left( {\dot r'\over \dot r r'} +{2\over r}\right) \approx 0
\, . \label{12-3}\end{equation} Now, let us find a solution to the
Einstein equations in which the function $y$ (as well as
$\varepsilon$ and $\nu$) depend explicitly only on the functions
$r$ and ${r_h}$  and do not depend explicitly on $t$ and $R$. When
the density of matter is defined so, the function ${\varepsilon
(r)}$ is determined solely by the initial distribution of the
density of matter.

As ${\delta\to 0}$, expressions (\ref{3-2-1}) and (\ref{9-1-2})
for $\nu_{_0}$ and $\lambda_{_0}$ must tend to ${\nu_{_\Lambda}}$
and ${\lambda_{_\Lambda}}$, respectively [see~(\ref{1-8})], and
$\alpha$ must tend to zero. Accordingly, we choose the energy
density distribution as
\begin{equation}\begin{array}{l}
\varepsilon = \varepsilon_{_\Lambda}\left[ 1-{\delta\over 2}
\ln\left( 1+\alpha- {r_h\over r} \right) \right] \, ,\\
\quad \\
\nu = -2{1+\delta\over\delta}y\approx (1+\delta)\ln\left(
1+\alpha- {r_h\over r} \right) \, .
\end{array}\label{12-5}\end{equation}
In the main approximation in $\delta$ with regard to the initial
conditions, we obtain
\begin{equation}
{\dot r' \over r'} \approx -{2\dot r\over
r}\quad\Longrightarrow\quad r' \approx \left( R\over r\right)^2 \,
. \label{12-4}\end{equation} Expression (\ref{9-3-1}) yields
\begin{equation}\begin{array}{l}
2M_{_0}\approx r_g +\Lambda R^3 - {3\over 2}\delta\Lambda r_{_1}^3
\cdot
I(R/r_{_1} -1) \, , \\
\quad \\
I(x)=x^3\left[{\ln x\over 3}-{1\over 9} \right] + x^2\left[{\ln
x\over 2}-{1\over 4} \right]+ x\left[\ln x-1 \right] -
(x+1)^3 \left[{\ln (x+1)\over 3}-{1\over 9} \right] -{1\over 9} \, , \\
\quad\\
\lim\limits_{x\to 0}I(x)=x(\ln x -1)\to 0 \, ,
\end{array}\label{12-7}\end{equation}
where ${r_{_1}=r_g +\Lambda r_{_1}^3\approx r_g}$.

Taking into account that $\varepsilon$ depends explicitly only on
$r$, we obtain the following expression from (\ref{9-7}):
\begin{equation}
\begin{array}{l}
2M = 2M_{_0}(R) - (1+\delta)\int\limits_r^R 8\pi\varepsilon r^2\,
dr = 2M_{_0}(r) -\delta \int\limits_r^R 8\pi\varepsilon r^2\, dr
\, .
\end{array}\label{12-6}\end{equation}
In the main approximation in $\delta$, formula (\ref{12-6}) is
rewritten as
\begin{equation}
2M\approx r_g +\Lambda r^3 -\delta\Lambda \left[ R^3-r^3 + {3\over
2}r_{_1}^3\cdot I(r/r_{_1}-1) \right] \, .
\label{12-8}\end{equation} From (\ref{3-3-0}) and (\ref{9-1-2}) we
find
\begin{equation}
e^\lambda \approx {(R/r)^4 \over 1+\alpha-S}\cdot
\left[{1-{\delta\over 2} \ln\left( 1+\alpha- r_h/r \right) \over
1-{\delta\over 2}\ln\left(1+\alpha- r_h/R \right) }
\right]^{2/\delta} \, . \label{12-10}\end{equation} Hence, using
(\ref{9-4}) and applying the second remarkable limit to
(\ref{12-10}), we obtain
\begin{equation}
e^\lambda\approx {(R/r)^4 \over 1+\alpha-S}\cdot \left({1+\alpha-
r_h/R \over 1+\alpha- r_h/r }\right) \, ,\quad V^2\approx
1-{1-2M/r \over 1+\alpha-S}\cdot \left({1+\alpha- r_h/R \over
1+\alpha- r_h/r }\right) \, . \label{12-11}\end{equation}
According to the second expression in (\ref{9-4}), the horizon
radius $r_h$ is determined by the formula ${r_h=2M}$. Then, from
(\ref{12-8}) we find that
\begin{equation}
r_h -r_{_1} \approx -\delta\Lambda (R^3-r_{_1}^3) \, .
\label{12-9}\end{equation} as ${\delta\to 0}$. However, the
solution obtained does not clearly indicate the direction of
evolution (either collapsing or expansion). The direction of
evolution is determined by the acceleration of matter at the
initial moment.

\section{ACCELERATION}
\label{s6}

Equations (\ref{1-3}) and (\ref{9-1-2}) yield the following
expression for the second derivative of radius $r$ with respect to
time $t$ at the initial moment:
\begin{equation}
2R e^{-\nu_{_0}}\ddot r_{_0}= (1+\delta) 8\pi\varepsilon R^2 +
R\dot r_{_0}\dot\nu_{_0} e^{-\nu_{_0}} -S + R {\nu'}_{_0}(1+\alpha
-S) \, . \label{2-1}\end{equation} At the initial moment, the
proper acceleration ${a=dV/d\tau}$ of the observer co-moving with
matter is given by (see~(\ref{6-5})):
\begin{equation}
a_{_0} = \ddot{r_{_0}}\cdot\exp (\lambda_{_0} /2 -\nu_{_0}) \, .
\label{2-2}\end{equation} For small $\delta$, from (\ref{12-5})
and (\ref{12-7}) we obtain
\begin{equation}
\begin{array}{l}
\nu_{_0}\approx (1+\delta) \ln\left( 1+\alpha -S \right) \, ,\\
\quad \\
S\approx {r_g\over R}+\Lambda R^2 - {3\over 2}\delta\Lambda
{r_{_1}^3\over R}
\cdot I(R/r_{_1}-1) \, , \\
\quad\\
RS'\approx -{r_g\over R}+2\Lambda R^2 + {3\over 2}\delta\Lambda
{r_{_1}^3\over R} \cdot \left[ I(R/r_{_1}-1)
-RI'(R/r_{_1}-1)\right] \, .
\end{array}\label{12-9}\end{equation}
Let us rewrite (\ref{2-1}) with regard to
(\ref{12-9})\footnote{Here, as before, the prime denotes
differentiation with respect to $R$.}:
\begin{equation}
2R e^{-\nu_{_0}}\ddot r_{_0} \approx \delta \left[{r_g\over
R}+\Lambda R^2 +{3\over 2}\Lambda r_{_1}^3 \cdot I' \right]
+\alpha {r_g/R - 2\Lambda R^2 + {3\over 2}\delta\Lambda r_{_1}^3
(I' -I/R) \over 1+\alpha -S} \, . \label{2-3}\end{equation} Hence,
we can see that the acceleration vanishes (as it must) at
${\delta=0}$ (and ${\alpha =0}$).

For radii $R$ close to ${r_{_1}=r_g+\Lambda r_{_1}^3}$, the terms
containing
$$
I' \approx \ln (R/r_{_1}-1)/r_{_1} <0 ,
$$
prove to be dominant in (\ref{2-3}). Denote
\begin{equation}
R_{cr}\approx r_{_1}\left[1+\exp\left( -{2\over 3\delta\Lambda
r_{_1}R_{cr}}\right)\right] \approx r_{_1}\left[1+\exp\left(
-{2\over 3\delta\Lambda r_{_1}^2}\right)\right] \approx r_{_1} \,
, \label{2-33}\end{equation} Then, from (\ref{2-3}) we find that,
for $\delta >0$, matter expands at the initial moment when
${R>R_{cr}}$ and collapses when ${(R<R_{cr})}$ (in the close
vicinity of the horizon).

\section{CHARACTERISTIC EVOLUTION TIMES OF THE SYSTEM}
\label{s7}

In this section, we will not assume that $\delta$ is small. Then,
neglecting $\alpha$ and using (\ref{2-1}), we can obtain the
following expression for the characteristic evolution time of the
system:
$$
T\sim\sqrt{r/ |\ddot r|}.
$$

\begin{figure}[t]
\centering \epsfbox[20 300 500 600]{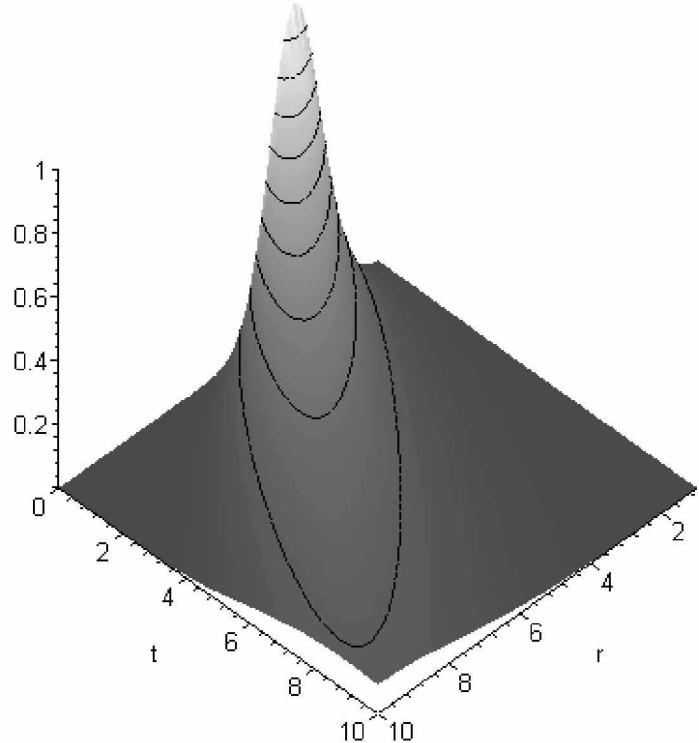}

\caption{The behavior of a cloud of phantom matter. The cloud,
which has the form of a Gaussian distribution when at rest, starts
do expand with time. The vertical axis represents the energy
density normalized to unity.}

\label{R2} \end{figure}

If the initial distribution of matter is characterized by large
gradients, one can neglect all the terms in (\ref{2-1}) except the
terms containing ${R{\nu_{_0}}'}$. Then, (\ref{2-1}) and
(\ref{2-2}) yield
\begin{equation}
\ddot r_{_0} \approx {w\over\delta} ({\varepsilon_{_0}}'
/\varepsilon_{_0}) (1-S) (\varepsilon_{_0}
/\varepsilon_{_\Lambda})^{2w/\delta}\, ,\quad a_{_0} \approx
{w\over\delta} ({\varepsilon_{_0}}' /\varepsilon_{_0})\sqrt{1-S}\,
. \label{2-5}\end{equation} According to formulas (\ref{2-5}),
there are two alternative versions of the evolution of matter with
the initial distribution in the form of a spherical layer with a
local maximum of energy density:

1. In the case of phantom matter ${(\delta >0)}$, we obtain a
expanding layer of matter (see Fig.~\ref{R2}).

2. When ${\delta <0}$, we obtain a collapsing layer of matter.

Note that the greater the density gradient, the faster the
evolution. Apparently, these results can be generalized to a
non-spherical model.

The characteristic evolution times corresponding to expressions
(\ref{2-5}) are given by
\begin{equation}
T_1\sim \sqrt{\left| (\delta /w) R \varepsilon_{_0}
(\varepsilon_{_\Lambda} / \varepsilon_{_0})^{2w/\delta}\right| /
\left| {\varepsilon_{_0}}' (1-S) \right| } \, ,\quad T_2\sim
\sqrt{\left|  (\delta /w) R \varepsilon_{_0} \right| /
\left|{\varepsilon_{_0}}' \sqrt{(1-S)} \right| } \, .
\label{2-6}\end{equation} Here, $T_1$ is the evolution time for an
infinitely distant observer, and $T_2$ is the evolution time for
an observer that moves with matter.

To conclude this section, we consider the evolution of the domain
that is characterized by the maximum energy density (a "hump"; see
Fig.~\ref{R2}). In this domain, ${R{\nu_{_0}}' \approx 0}$;
therefore, formulas (\ref{2-1}) and (\ref{2-2}) are rewritten in
this domain as
\begin{equation}
\begin{array}{l}
\ddot r_{_0}\approx \exp (\nu_{_0}) \left[(1+\delta) \cdot
8\pi\varepsilon_{_0} R^2 -S
\right]/(2R) \, ,\\
a_{_0} \approx \left[(1+\delta) \cdot 8\pi\varepsilon_{_0} R^2 -S
\right]/[2R\sqrt{1-S}]\, .
\end{array}\label{2-7}\end{equation}
For the initial distribution of matter similar to that shown in
Fig.~\ref{R2}, we have
$$
S=\frac{\int\limits_{_{_0}}^{_R}8\pi\varepsilon R^2\, dR}{R} <
8\pi\varepsilon_{_0} R^2,
$$
at the maximum; therefore, the acceleration at this point proves
to be positive; i.e., the "hump"\, decomposes and flies away from
the center of the system (for any ${\delta >0}$).

\section{DYNAMICS OF THE APPARENT HORIZON}
\label{s8}

Since all particles may intersect the horizon only in one
direction, a strict correspondence is established between the
coordinates ${R_h}$ and ${t_h}$ on the horizon. This means that
${R_h(t)}$ is a continuous and monotonically increasing function
${(\dot R_h >0)}$.

Consider formula (\ref{9-7}) on the horizon. Differentiating this
formula with respect to time, we obtain the following expression
for the variation of the horizon radius:
\begin{equation}
\dot r_h=8\pi\varepsilon_{_0} R_h^2\cdot {dR_h\over dt} - 8\pi
w\varepsilon r^2\dot r \, . \label{7-2} \end{equation} Taking into
account (\ref{6-1-2}), we rewrite (\ref{7-2}) as
\begin{equation}
\dot r_h=8\pi\varepsilon_{_0} R_h^2\cdot \dot R_h \cdot \left( 1 +
wr'\cdot {\varepsilon\over\varepsilon_{_0}}\cdot{r^2\over R^2}
\right) \label{7-2-1} \end{equation} Consider two ways in which
the horizon radius varies.

\subsection{Test Matter}
\label{test}

Consider the dynamics of the horizon for test matter (which
negligibly affects the variations of the system and the radius of
the horizon). For such matter, all functions must depend only on
$r$. Otherwise, the metric in a static (fixed) frame of reference
turns out to be nonstatic, which implies that it is affected by
matter (i.e., the assumption that it is a test matter is
violated).

Equation (\ref{6-1}) determines the constancy of $r$, while the
constancy of the metric component $\nu$ is given by
\begin{equation}
d\nu = 0= \nu' \, dR + \dot\nu \, dt  \, .
\label{7-3}\end{equation} Solving Eqs. (\ref{6-1}) and (\ref{7-3})
simultaneously, we obtain
\begin{equation}
\nu' \dot r = \dot\nu r'  \, . \label{7-4}\end{equation} Equations
(\ref{1-5-2}) and (\ref{7-4}) yield
\begin{equation}
2\dot{r'} /r' =\dot\lambda + \dot\nu \, .
\label{7-5}\end{equation} Integrating this equation with regard to
the initial conditions, we obtain
\begin{equation}
r'=\sqrt{\exp{(\lambda -\lambda_{_0} +\nu -\nu_{_0})}} \, .
\label{7-6}\end{equation} Hence, taking into account (\ref{3-3-0})
and (\ref{3-2-1}), we find
\begin{equation}
r'={\varepsilon_{_0}\over\varepsilon}\cdot{R^2\over r^2} \, .
\label{7-8}\end{equation} Now, formula (\ref{7-2-1}) for the
variation of the horizon radius is rewritten as
\begin{equation}
\dot r_h = 8\pi\varepsilon_{_0} R_h^2\cdot \dot R_h \cdot (1+w) \,
. \label{7-9} \end{equation} Thus, for ${w<-1}$, we have ${\dot
r_h<0}$; i.e., in the case of test matter, the horizon radius
decreases, and the black hole is "dissolved"\, in phantom matter.

In formula (\ref{7-9}), the fact that matter is test manifests
itself in that the quantity ${8\pi\varepsilon_{_0} R^2}$ is
extremely small, ${8\pi\varepsilon_{_0} R^2 <<<1}$. This result
was predicted and proved for the static frame of reference (a
frame of reference fixed with respect to distant stars)
in~\cite{3}.

\subsection{Nontest Matter}
\label{notest}

Formula (\ref{7-9}) can be generalized to nontest matter. To this
end, it suffices to notice that, when the horizon radius is
constant, the metric coefficients on the horizon are also
time-independent (the gravitational field at radius $r$ is
determined only by the mass within this radius, i.e., by the mass
within the horizon radius in the case considered). Therefore, the
following relation must hold for ${\dot r_h=0}$:
$$
d\nu |_{r_h}=0 .
$$
Using this in expression (\ref{7-3}) on the horizon, we obtain the
same result (\ref{7-9}). Hence, if the accretion of matter does
not change the horizon radius, then the equality ${w = -1}$ must
hold.
\begin{equation}
\left. d\nu\right|_{r_h}= \dot\nu\, dt + \nu' \, dR \, .
\label{7-11} \end{equation} Solving Eqs. (\ref{6-1}) and
(\ref{7-11}) simultaneously, we obtain
\begin{equation}
\nu' \dot r = \dot\nu r' - r'\, \left. {d\nu\over
dt}\right|_{r=r_h=\const} \, . \label{7-12} \end{equation} which
is an analog of Eq. (\ref{7-4}). Equations (\ref{7-12}) and
(\ref{1-5-2}) yield
\begin{equation}
2\dot{r'} /r' =\dot\lambda + \dot\nu - \left. {d\nu\over
dt}\right|_{r=r_h=\const} \, . \label{7-13}\end{equation}
Integration of this equation with the initial conditions yields
analog of Eq. (\ref{7-6}):
\begin{equation}
r'=\sqrt{\exp (\lambda -\lambda_{_0} +\nu -\nu_{_0}) \cdot \exp
\left( \nu_{[1]} - \nu_{[2]} \right) } \, ,
\label{7-14}\end{equation} Here, the following notation is used:
${[1]\equiv (r_h,0)}$ and ${[2]\equiv (R_h,t_h)}$;
${r_{[1]}=r_{[2]}=r_h}$. An expression for the variation of the
horizon radius is obtained analogously (\ref{7-9}):
\begin{equation}
\dot r_h = 8\pi\varepsilon_{_0} R_h^2\cdot \dot R_h \cdot
[1+w\Psi]\, , \quad \Psi\equiv \sqrt{\exp \left( \nu_{[1]} -
\nu_{[2]} \right)}=
\left({\varepsilon_{[1]}\over\varepsilon_{[2]}}\right)^{w/\delta}
\, . \label{7-15} \end{equation} These are exact expression
independent of $\delta$. Thus, the direction in which the horizon
is shifted depends on the initial conditions of the distribution
of matter. The condition under which the horizon radius decreases
is given by ${w\Psi<-1}$.

For a matter distribution that depends explicitly only on $r$, we
obtain ${\Psi =1}$. Then, formulas (\ref{6-2}), (\ref{12-5}),
(\ref{12-11}) and (\ref{7-15}) yield
\begin{equation}
\dot r_h \approx - 3\left({\delta\over\sqrt{\alpha}}\right)
\Lambda r_h^2 (1+\alpha-S)^{3/2} \quad\Longrightarrow\quad
T_{r_{_1}\to r_h}\approx {\sqrt{\alpha}(r_{_1} - r_h)\over
3|\delta|\Lambda r_{_1} r_h (1+\alpha-S)^{3/2}} \, .
 \label{7-16} \end{equation}
This formula shows that, in the general case, the "dissolution"\,
degree of a black hole for small $\delta$ is rather small.

\section{CONCLUSIONS}
\label{s9}

The main conclusion of the present study is the confirmation of
the conjecture on a possible decrease of the horizon radius
(solution) of a black hole.

Note that, at infinity, there is no paradox related to this
phenomenon. In the absence of matter, the metric at infinity
persists to be Schwarzschild (with a $\Lambda$-term). Accordingly,
the total mass of the whole system related to this metric remains
constant. The point is that the mass and energy are redistributed
in space during evolution: an external wave of phantom matter
removes the mass and energy that correspond to the initial black
hole (which is "dissolved"\, under the action of the internal wave
of phantom matter). If the black hole is not "dissolved"\,
completely, the total energy of the system is composed of the mass
of the changed black hole and the energy of the external wave of
phantom matter.

$\quad$

{\bf ACKNOWLEDGMENTS: \label{s11}} \\
I am grateful to I.D. Novikov and N.S. Kardashev for discussing
various questions while preparing this article.

%\newpage

\end{document}